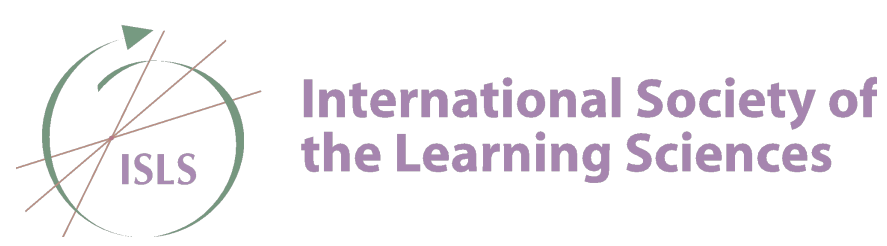

# Modeling AI-TPACK in Practice: Insights from Teachers' Multi-Agent Workflow Design

Yimeng Sun, Haiyang Xin, Shuang Li, Qiannan Niu
sunyimeng@cocorobo.cc, Tony@cocorobo.cc, lishuang@cocorobo.cc, niuqiannan@cocorobo.cc,
CocoRobo LTD
Ching Sing Chai, The Chinese University of Hong Kong, CSChai@cuhk.edu.hk
Lingyun Huang, The Education University of Hong Kong, lingyunhuang@eduhk.hk
Gaowei Chen, The University of Hong Kong, gwchen@hku.hk

**Abstract:** This study investigates teachers' design behaviors and cognitive underpinnings when designing multi-agent instructional workflows. Analyzing behavioral logs (N=61), cluster and Markov analyses identified three archetypes: *Systematic Optimizers* iteratively refining complex architectures; *Prolific Creators* rapidly prototyping pragmatic tools via scaffolding; and *Passive Observers* exhibiting polarized expert-novice profiles. Subsequent artifact (n=15) and interview (n=12) analyses reveal AI-TPACK integration emerges from a dynamic interplay of systems thinking, pedagogical beliefs, and self-efficacy, not merely from the possession of discrete knowledge. These findings call for differentiated scaffolding responsive to teachers' cognitive-behavioral diversity.

## Introduction

The maturation of large language models (LLMs) offers transformative potential for education by enabling personalized learning support and interactive pedagogical innovations (Kasneci et al., 2023; Zhai, 2022). No-code platforms enabling multi-agent system design are democratizing this capability (Potharalanka, 2025; Özdemir et al., 2025), shifting teachers from AI consumers to AI designers. However, this shift demands teachers master "architectural thinking", designing and coordinating complex agent logic for pedagogical goals. Traditional TPACK (Mishra & Koehler, 2006)—a framework describing teachers' integration of technological, pedagogical, and content knowledge—inadequately addresses this competency, as AI agents' generative and conversational nature fundamentally differs from prior educational technologies. The AI-TPACK framework (Celik, 2023; Ning et al., 2024) addresses this gap by extending TPACK with domains critical for AI integration, such as understanding AI affordances, prompt engineering, and leveraging AI to enact pedagogical strategies.

Despite the importance of this new competency, research on teachers' actual AI-TPACK integration remains limited in three key ways. First, studies predominantly rely on self-report surveys (e.g., Karataş & Ataç, 2025), which capture stated beliefs rather than actual design practices. Objective, process-oriented data (e.g., design logs) are needed to understand teachers' "in-the-moment" design strategies (Hadwin et al., 2007). Second, research often treats teachers as a homogenous group, masking diverse behavioral profiles and integration patterns. Third, we lack explanatory models linking teachers' observable design behaviors and artifact-level outcomes to their underlying cognitive-affective factors. It remains unknown what behavioral archetypes of AI workflow design are, how they map to AI-TPACK quality, or why they emerge. Answering these questions can advance AI-TPACK theory through empirical grounding while informing differentiated professional development. This study investigates teachers' design practices during a two-day AI professional development workshop, guided by the following research questions:

- RQ1: What behavioral patterns do teachers exhibit when designing multi-agent instructional workflows?
- RQ2: How do teachers with different behavioral patterns differ in their AI-TPACK manifestations as evidenced in design artifacts?
- RQ3: What cognitive, affective, and competency factors explain these differences in teachers' AI-TPACK integration?

## Methods

### Context and Participants

This study involved 61 K-12 in-service teachers (34 female, 27 male; teaching experience: 1-28 years, M=8.07, SD=7.36) from diverse disciplines in a coastal city in Southern China. They voluntarily participated in a two-day district-wide AI professional development workshop, aimed at designing discipline-specific AI-powered learning tools. By the end, each teacher had produced one functional multi-agent workflow.

## Data Collection

Teachers used CocoFlow, a no-code multi-agent development platform designed according to 'low floor and wide walls' principles (Resnick & Silverman, 2005). Each teacher designed one workflow by connecting multiple specialized AI agents (e.g., content tutor, feedback provider) with functional tool nodes (e.g., forms for data collection, OCR for image recognition) through sequential, conditional, or parallel logic. Teachers configured each agent's behavior via natural language prompts, shifting focus from technical implementation to pedagogical logic. CocoFlow embedded analytics system captured comprehensive process-oriented behavioral data, including feature usage patterns (e.g., configuration, testing, browsing), temporal sequences, and design iterations. These clickstream data (N=61 teachers, 8,718 actions) served as objective indicators of teachers' design strategies (Hadwin et al., 2007; Siemens, 2013).

## Data Analysis

We employed a mixed-methods explanatory sequential design (Ivankova et al., 2006). First (RQ1), platform log data were analyzed using K-means clustering in R to identify behavioral profiles based on action type distributions. The number of clusters ($\kappa$=3) was determined via the elbow method, yielding an average silhouette score of 0.389. Markov transition matrices characterized each cluster's workflow dynamics.

Second (RQ2 & RQ3), the qualitative phase was guided by the AI-TPACK framework, as specified for the generative AI context by Ning et al. (2024). This model extends the traditional TPACK by articulating the complex, multi-dimensional knowledge teachers require for AI integration. We operationalized this framework as our a priori coding scheme, focusing on the four key AI-specific dimensions: AI-Technological Knowledge (AI-TK), AI-Technological Pedagogical Knowledge (AI-TPK), AI-Technological Content Knowledge (AI-TCK), and the overarching AI-TPACK (Integration).

Using this framework, we conducted a two-stage thematic analysis (Braun & Clarke, 2006). For RQ2, design artifacts (n=15: 5 per cluster, purposively sampled for representativeness) were analyzed. Two researchers independently coded these workflows across the four AI-TPACK dimensions, achieving acceptable inter-rater reliability (Cohen's κ=0.81). For RQ3, semi-structured interviews with 12 volunteers, purposively sampled across all three behavioral clusters (C1: n=4; C2: n=5; C3: n=3), explored their AI-TPACK cognition, pedagogical beliefs, and self-efficacy. Six of these participants were also included in the artifact sample, allowing direct linking of cognitive profiles and design outputs. Interview transcripts were thematically analyzed by the same two researchers using the same four-dimension framework, with coding consistency maintained through iterative independent coding and consensus discussion. Finally, codes from artifacts (RQ2) and interviews (RQ3) were synthesized with behavioral findings (RQ1) to construct explanatory profiles linking cognition to practice.

# Results

## RQ1: Teachers' Behavioral Patterns in Multi-Agent Workflow Design

To understand what behavioral patterns teachers exhibit when designing multi-agent instructional workflows, we conducted a two-stage analysis combining K-means cluster analysis of platform action distributions with Markov transition analysis of action sequences. This approach enabled us to characterize both the composition and dynamic structure of teachers' design processes.

K-means clustering ($\kappa = 3$) based on teachers' proportional engagement across eight action categories revealed three distinct behavioral profiles. Cluster 1 (n=18) was configuration-intensive: Content Configuration dominated their activity (32.50%), while Content Editing and Testing/Debugging rates were 3-4 times higher than other clusters. Cluster 2 (n=35) balanced creation and browsing: Object Creation (25.20%) and Browsing Own Content (27.10%) comprised over half their activity, but editing and testing remained minimal. Cluster 3 (n=8) was browsing-dominant: Browsing Own Content (29.00%) was their primary activity, with the lowest editing rate (1.70%) and highest Template Cloning (2.45%) across clusters.

Markov transition analysis revealed how these compositional differences manifested in workflow structures. Cluster 1 exhibited iterative optimization loops: bi-directional transitions between Testing and Configuration, combined with Editing-to-Configuration pathways, formed a tightly coupled refinement cycle. High Configuration self-loops (0.732) but moderate Testing self-loops (0.351) indicated sustained parameter adjustment punctuated by rapid evaluation-adjustment pivots. Cluster 2 demonstrated rapid prototyping flows: Template Cloning transitioned to Object Creation at nearly twice the rate of other clusters (0.684), followed by sustained Creation self-loops (0.458), suggesting efficient conversion of templates into continuous creative output. Cluster 3 revealed browsing-anchored exploration: their most frequent transition was Browsing Others to Browsing Own (0.366), forming a self-reinforcing observation loop. Uniquely, they showed Template Cloning self-loops (0.482) and frequent Creation-to-Browsing transitions (0.278), indicating hesitant creation followed by immediate validation-seeking rather than sustained production.

These patterns revealed three archetypes: Systematic Optimizers (iterative refinement), Prolific Creators (template-based prototyping), and Passive Observers (browsing-anchored participation). Having established these behavioral profiles, we next examine how they manifest in teachers' design artifacts through the AI-TPACK framework (RQ2).

## RQ2: AI-TPACK Manifestations in Multi-Agent Workflow Designs

Having established behavioral profiles, we examined how these patterns manifested in teachers' design artifacts through AI-TPACK-informed thematic analysis of workflows (n=15: 5 per cluster). Rather than describing each cluster comprehensively, we organize findings by the AI-specific dimensions of our framework (AI-TK, AI-TPK, AI-TCK, and AI-TPACK) to reveal systematic variations (Table 1).

**Table 1**

*AI-TPACK Manifestations in Workflow Design by Cluster*

| Dimension | C1: Optimizers | C2: Creators | C3: Observers |
|---|---|---|---|
| AI-TK | Complex multi-agent logic; iterative refinement | Template-based configurations; moderate complexity | Polarized: detailed prompts vs. config-only designs |
| AI-TPK | Systemic tutoring systems; sustained guidance | Modular problem-solving; contextualized engagement | Differentiated instruction vs. basic interaction |
| AI-TCK | Deep feedback (essay grading); scaffolded learning | Targeted knowledge point support (equations, syntax analysis) | Concept encoding vs. minimal content customization |
| AI-TPACK | High: cohesive design across dimensions | Pragmatic: functional enhancement of teaching | Unstable: expert-level vs. surface-level integration |

### AI-Technological Knowledge (AI-TK): From Functional Use to Architectural Mastery

Teachers exhibited a spectrum of AI-TK sophistication. At the functional end, Cluster 2 teachers leveraged accessible platform features to construct straightforward agent configurations, where designs like "Equation Solver Assistant" delivered targeted support without complex logic. In contrast, Cluster 1 teachers demonstrated architectural thinking: their "Essay Grading System" coordinated four agents: a Content Analyzer (evaluating arguments), Language Reviewer (assessing mechanics), Feedback Synthesizer (integrating analyses), and Revision Tutor (providing scaffolded guidance for low-scoring essays via conditional branching). One teacher articulated this challenge as "coordinating logic between agents," revealing conscious engagement with multi-agent integration. Notably, they pushed beyond platform constraints, requesting capabilities like "image recognition", indicating awareness of AI's potential exceeding current implementations. Cluster 3 bifurcated dramatically: while 2 of 5 teachers produced workflows with detailed instructions encoding precise logic, the other 3 of 5 created designs with no prompts, relying entirely on defaults. This AI-TK variance foreshadowed integration depth differences.

### AI-Technological Pedagogical Knowledge (AI-TPK): Pedagogical Vision and Pragmatism

Pedagogical orientations diverged sharply. Cluster 1 positioned AI as systemic learning companions providing continuous scaffolding. Their "AI companion" workflows enacted sustained, personalized guidance aligned with constructivist principles. Cluster 2 adopted instrumental pedagogy: AI amplified specific teaching strategies (e.g., "Science Stories" embedded concepts in narrative to boost engagement) rather than reimagining learning systems. Their AI-TPK was modular and contextualized, targeting immediate instructional pain points. Cluster 3's high-capability subset designed differentiated pathways adapting to learner responses, while low-capability members produced generic interaction with minimal pedagogical customization. Critically, AI-TPK sophistication aligned with behavioral patterns (RQ1): systemic visions required iterative refinement (C1), while instrumental approaches enabled rapid prototyping (C2).

### AI-Technological Content Knowledge (AI-TCK): Disciplinary Integration Depth

All clusters targeted disciplinary challenges, but integration depth varied. Cluster 1 achieved deep encoding: they uploaded discipline-specific knowledge-base files (e.g., detailed rubrics, disciplinary writing standards) and translated this expertise into sophisticated, multi-dimensional feedback mechanisms within the agent instructions. Cluster 2 demonstrated targeted application: "Sentence Structure Analyzer" and "mathematical problem-solvers" addressed specific knowledge gaps with efficient, scalable support, but lacked the architectural depth to support complex disciplinary reasoning. Cluster 3's AI-TCK ranged from precise conceptual progression (high-capability subset: "Understanding AI" workflow) to absent customization (low-capability subset: content-agnostic configurations).

#### AI-TPACK (Integration): Coherence Across Dimensions

Integration, the synergy of AI-TK, AI-TPK, and AI-TCK, distinguished clusters most clearly. Cluster 1 achieved unified systems where technical architecture served pedagogical intent grounded in disciplinary expertise: complex agents, sustained guidance, and deep content encoding cohered into systemic designs. Cluster 2 displayed pragmatic enhancement: technology amplified existing practices, with integration depth contingent on external scaffolding: supported teachers produced sophisticated workflows, while independent designers created simpler feature-stacking designs (i.e., workflows that piled on platform functions without a coherent pedagogical or technical integration). Cluster 3 exhibited unstable integration: expert-level coherence coexisted with surface-level configurations lacking pedagogical or content specificity. This polarization suggested behavioral patterns alone insufficiently predict AI-TPACK capacity; cognitive-affective factors (RQ3) must be considered.

### RQ3: Cognitive, Affective, and Competency Underpinnings of AI-TPACK Differences

Interview analysis revealed that behavioral patterns were underpinned by systematically different cognitive and affective profiles. Each cluster demonstrated distinct levels of AI-TK, technical self-efficacy, and pedagogical beliefs explaining their design behaviors and outcomes.

Systematic Optimizers (Cluster 1) demonstrated sophisticated AI-TK centered on multi-agent integration. One teacher explained: "The hardest part is the logical relationships between agents. I designed a diagnostic agent, tutoring agent, and practice agent, but how to make them collaborate—when to jump to the next agent—this control took me a long time to adjust." This systems-level thinking explained their iterative configuration-testing cycles (RQ1) and architecturally complex workflows (RQ2). Their AI-TPK embodied constructivist principles, positioning AI as persistent learning companions. They exhibited high technical self-efficacy while externalizing barriers to platform limitations, sustaining high-investment optimization despite challenges.

Prolific Creators (Cluster 2) revealed contrasting profiles. Their AI-TK remained at functional understanding without architectural depth. One stated: "we're quite weak in technology"; another noted: "barely passing...my ability is far from sufficient". This low self-efficacy explained their modular, feature-stacking designs (RQ2). Their AI-TPK was instrumentally pragmatic: AI alleviated grading burdens and targeted knowledge gaps. Most critically, they exhibited extreme support-dependence, requiring continuous on-site guidance. This dependency explained their template-seeded rapid prototyping (RQ1), where templates provided entry points while external scaffolding bridged capability gaps.

Passive Observers (Cluster 3) displayed polarized cognition mirroring their design dichotomy (RQ2). High-capability teachers critiqued platform constraints like "lacking open API access", while low-capability teachers felt "completely clueless" despite platform accessibility. Both shared positive attitudes toward AI yet extreme support-dependence, explaining their browsing-anchored hesitation (RQ1): observation substituted for autonomous experimentation.

## Discussion

Our findings demonstrate that AI-TPACK integration is a dynamic process shaped by self-efficacy and support structures, not merely discrete knowledge possession. AI-TK in multi-agent contexts demands systems thinking beyond operational proficiency—a competency absent from traditional TPACK. AI-TPK differentiates between systemic reimagination (C1) and instrumental enhancement (C2), suggesting AI enables fundamentally different pedagogical philosophies.

These insights yield actionable design principles. First, differentiated scaffolding is essential: Cluster 3 requires structured templates and real-time assistance to overcome low self-efficacy; Cluster 2 benefits from pedagogical frameworks and exemplar repositories bridging capability gaps; Cluster 1 needs advanced documentation and open architectural affordances. Second, platforms should implement adaptive support systems using behavioral analytics to dynamically identify teacher profiles and deliver personalized resources. Third, trajectories from novice to expert require progressive scaffold fading: platforms must design pathways helping teachers transition from template replication (C3) through pragmatic application (C2) toward systemic innovation (C1), ensuring support-dependence evolves into autonomous mastery.

While these profiles offer clear implications, findings are situated within a short-term professional development context requiring further validation. Supporting teachers' transition from AI consumers to AI designers requires recognizing AI-TPACK development as a differentiated process demanding adaptive scaffolding.

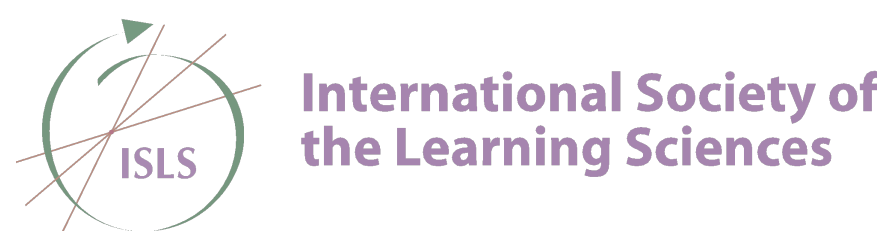